%% file: main.tex
\documentclass[useAMS,usenatbib]{mn2e}
\usepackage{graphicx}
\usepackage{color}
\usepackage{url}
\input{newcom.tex}

\newcommand{\he}{HEAsoft}
\newcommand{\xspec}{\emph{Xspec}}

\newcommand{\compps}{{\sc compps}}

\newcommand{\totalmod}{{\sc const(phabs(compps+gauss))}}

\usepackage{amsmath}
\usepackage{amssymb}
\usepackage{float}

\title{The impact of neutron star spin on X-ray spectra}

\author[M. J. Burke, M. Gilfanov  \& R. Sunyaev]{M. J. Burke$^{1}$\thanks{E-mail:
mburke@mpa-garching.mpg.de (MJB);}, M. Gilfanov$^{1,2,3}$ and R. Sunyaev$^{1,2}$    \\
$^{1}$ Max Planck Institute for Astrophysics, Karl-Schwarzschild-Str. 1, Garching b. Munchen D-85741, Germany\\
$^{2}$ Space Research Institute of Russian Academy of Sciences, Profsoyuznaya 84/32, 117997 Moscow, Russia \\
$^{3}$ Kazan Federal University, Kremlevskaya str.18, 420008 Kazan, Russia}

\begin{document}

\date{Accepted Year Month Day. Received Year Month Day; in original form Year Month Day}

\pagerange{\pageref{firstpage}--\pageref{lastpage}} \pubyear{2014}

\maketitle

\label{firstpage}

\begin{abstract}
{We investigate whether the intrinsic spin of neutron stars leaves an observable imprint on the spectral properties of X-ray binaries.  To evaluate this we consider a sample of nine NSs for which the spins have been measured that are not accreting pulsars (for which the accretion geometry will be different).  For each source, we perform spectroscopy on a majority of RXTE hard state observations. Our sample of sources and observations spans the range of the Eddington ratios $L_X/L_{Edd}\sim0.005-0.100$. }
  
We find a clear trend between key Comptonization properties and the NS spin for a given accretion rate.  Specifically, at a given $L/L_{Edd}$, for more rapidly rotating NSs we find lower seed photon temperatures  and a general increase in Comptonization strength, as parametrised by the Comptonization $y$~parameter and amplification factor $A$.  This is in good agreement with the theoretical scenario whereby less energy is liberated in a boundary layer for more rapidly spinning NSs, resulting in a lower seed photon luminosity and, consequently, less Compton cooling in the  corona.   This effect in extremis results in the hard states of the most rapidly spinning sources encroaching upon the regime of Comptonization properties occupied by BHs.

\end{abstract}


\section{Introduction}

X-ray binaries (XBs) possess rapidly spinning neutron stars (NSs) that are `spun up' by accretion, an idea that was initially motivated by the need to identify possible progenitors { \citep{1976ApJ...207..574S,1976SvAL....2..130B}} for pulsars with rotation periods in the millisecond regime \citep{1975ApJ...195L..51H}. The large collecting area and excellent timing capabilities of the RXTE spacecraft \citep{1993A&AS...97..355B}  led to a variety of high frequency phenomena being discovered at X-ray energies, such as accreting millisecond pulsars \citep{1998Natur.394..344W}.  Another early breakthrough saw the detection of millisecond oscillations during type-I X-ray bursts \citep{1996ApJ...469L...9S}, which are characterised by a rapid increase in the X-ray flux over a period of $1-10$ seconds that subsequently decays over $0.1-10$ minutes.  Such bursts are the result of unstable thermonuclear burning of accreted material on the NS surface \citep{1995ApJ...438..852B,2006csxs.book..113S}. While the exact mechanism of the burst oscillations is still open for debate \citep[see][for review]{2012ARA&A..50..609W}, repeated observations over months and years show consistent frequencies and behaviour for those of any particular source \citep[see][for review of burst properties]{2008ApJS..179..360G}. Consensus formed that the oscillations observed during bursts are the result of some anisotropy in the emission from the NS surface, an interpretation that was vindicated by discovery of burst oscillations from millisecond pulsars with frequencies within $6\times10^{-3}$~Hz of the known rotation frequencies \citep{2003Natur.424...42C,2003ApJ...596L..67S}.  A sound understanding of NS spin and its effects will play a crucial role refining the NS equation of state \citep[see overview by][]{2016EPJA...52...63M}, and has enormous potential for the emerging field of gravitational wave astrophysics \citep[e.g.][]{1998ApJ...501L..89B,2005MNRAS.361.1153A,2015PhRvD..92b3006M}. 

In our previous work \citep[][hereafter Paper~I]{2017MNRAS.466..194B} we reported a possible connection between hard state spectral properties of NS XBs and their spin.  This result was not conclusive, but pointed towards a simple picture of the effect of spin on the spectral shape.  We found hints of a positive correlation between spin and the strength of Comptonisation, which we characterised in terms of the amplification $A$, the ratio of the total luminosity to the seed photon luminosity, and the Compton $y$-parameter, which is a measure of the average change in energy experienced by a population of photons travelling through a finite medium.  Our interpretation is that accreting material surrenders more energy as it impacts upon less rapidly rotating NSs which in turn produces more soft seed photons for Comptonisation leading to increased Comptonization losses for hot elections,  as proposed by \citet{1988AdSpR...8..135S} and  \citet{1989ESASP.296..627S}.

The main result of Paper~I is a dichotomy observed in the Comptonisation strength between hard state spectra of NS and BH XBs.  Comptonisation in NSs is suppressed by additional seed photons that are produced in the vicinity of the NS surface, on the surface itself or in a boundary layer between the surface and geometrically thin accretion disc.   Theoretically such a boundary layer forms to facilitate the change in angular momentum between the innermost portions of the disc and the surface \citep{1988AdSpR...8..135S} giving rise to blackbody-like emission, which has been observed using Fourier-resolved spectroscopy of soft state spectra \citep{2003A&A...410..217G,2006A&A...453..253R}.   It is {not clear whether } the classical boundary layer exists during the hard state  \citep[see][for review]{2007A&ARv..15....1D,2010LNP...794...17G} when the innermost portion of the accretion disc is thought to be truncated at many gravitational radii, however, there must still be some mechanism to bridge the difference in linear velocities { between the accreting material and the neutron star surface}. {  When the disc extends to the surface of the NS, at moderate accretion rates ($L/L_{Edd}\sim 0.1$), the scale height of the boundary layer is expected to be $\sim 1-2$~km, whereas at high $L/L_{Edd}$ the scale height expands such that the NS is enveloped by accreting material \citep[see][fig.~1]{2001ApJ...547..355P}. However, in this work we consider only sources in the hard spectral state which roughly correspond to luminosities of $L/L_{Edd}\lesssim 0.1$. }

{ In the Newtonian approximation the boundary layer contributes approximately 50\% of the total source emission \citep{1988AdSpR...8..135S}. However, \citet{1998AstL...24..774S} \& \citet{2000AstL...26..699S} showed that by considering realistic space time metrics the energy released by the boundary layer for a non-rotating NS is of order $70\%$ of the total emission, and that this fraction decreases with increasing frequency of prograde rotation.  In Paper~1 we estimated the emission from the vicinity of the NS as a fraction of the total emission to be in the range $\approx \frac{2}{3}-\frac{1}{2}$. }

There are few concrete connections between spin frequencies and other observable quantities.   Possible connections between the { NS spin}   and the frequencies of the so-called kHz Quasi-periodic oscillations (QPOs) remain the most studied and debated proposition.  kHz QPOs \citep{1996ApJ...469L...1V} are high frequency QPOs that are occasionally detected in pairs that vary in a correlated way \citep{2005MNRAS.357.1288B} by tens to hundreds of Hz \citep[see][for review of X-ray variability]{2006csxs.book...39V}. Early studies noted that the measured difference in frequencies of the two kHz QPO was compatible with either the NS spin frequency or its half, depending on the source \citep{2000ARA&A..38..717V, 2001AdSpR..28..481P}.  This scenario is attractive in the context of beat frequency models, where the lower frequency QPO arises from the beat between the upper kHz QPO and spin frequency \citep{1998ApJ...508..791M,2001ApJ...554.1210L}. However, other authors have argued that QPOs are actually a manifestation of the effects of strong gravity \citep[e.g.][]{1999PhRvL..82...17S,2003A&A...404L..21A,2004A&A...423..401Z,2008A&A...489..963S}. \cite{2014AN....335..168W} plot 12 sources with measured kHz QPO pairs against spin, and conclude there is a `clustering with a high scatter'  around $\Delta \nu_{kHz} \approx 0.5 \nu_{spin}$ and $\Delta \nu_{kHz} \approx \nu_{spin}$, but it is not clear whether there is a relationship. {It was also found that  the kHz-QPO frequency correlates with the X-ray flux and spectral shape as characterised by the position on the colour-colour diagram \citep{1999ApJ...511L..49M,2001AdSpR..28..307B}}.

\cite{2001ApJ...553L.157M} were the first to suggest different patterns of behaviour between `slow' ($\approx 300~$Hz) and `fast' ($\approx 600~$Hz) sources, noting that burst oscillations were almost always found during bursts exhibiting PRE for fast sources, whereas slower sources showed oscillations in all types of burst.  In subsequent studies the picture proved to be more complicated; bursts with and without PRE  are found from both slow and fast samples, but predominantly bursts containing oscillations are PRE bursts in the most rapidly rotating systems, and non-PRE bursts for other sources \citep{2004ApJ...608..930M,2008ApJS..179..360G}.  In addition to this,  \cite{2008ApJS..179..360G} found that burst durations are consistently short for slow sources, while the bursts for faster sources tend to have longer durations. \cite{2007ApJ...663.1252P} suggest that predominantly shorter bursts from more slowly rotating NSs could be a result of the greater velocity differential between the NS surface and the infalling material leading to greater mixing between the ashes of burned material and that which is freshly accreted, exhausting H and leading to faster, H-rich bursts.  On the other hand, slow rotators could be accreting from companions that happen to be degenerate, which is why the burning material is lacking in H, and if confirmed would indicate a different evolutionary path for differently spinning  sources \citep{2008ApJS..179..360G}.

{ In this paper we set out to identify possible links between the persistent hard state spectra of NS XBs and their spin.  In doing so we build upon the work of Paper~I, first expanding our sample of NSs of known spin and the number of spectra for each source, and then by comparing their behaviour as a function of accretion rate. 
It may be worth noting that the effects we report here are unrelated to the GR effect of the NS spin similar to the effect of the BH spin on X-ray spectral formation in the vicinity of Kerr black hole. Rather, these effects are a result of the dependence of the energy released in the boundary layer on the velocity difference between the accreting material and the NS surface. Although accurate quantitative prediction of these effects requires knowledge of the precise metrics of the space-time in the vicinity of the neutron star \citep{1998AstL...24..774S}, they can be understood in terms of Newtonian mechanics through the well known formula

\begin{equation}
L_{bl}=\frac{1}{2}\,\dot{M} \left(\alpha v_K- v_{NS}\right)^2
\label{eq:lbl}
\end{equation}
where $\alpha v_K$ is  velocity of the in falling material expressed in terms of the Keplerian velocity $v_K$ near the NS surface, the factor $\alpha~(\le 1)$  accounts for the possibility that in the hard state the accreting material  can arrive at the stellar surface with sub-Keplerian velocity, and $v_{NS}$ is linear velocity of the NS surface \citep{1988AdSpR...8..135S, 1988PhDT.........2K}}.

\section{Sample Selection}
\label{sec:sample}
To add additional sources to our sample we choose to consider only those for which the NS spin frequency is well determined.  This limits our considerations to only those sources for which oscillations have been observed during thermonuclear bursts \citep[see][]{2008ApJS..179..360G}, which we further refine by excluding millisecond pulsars, as these will almost certainly obey a different accretion geometry from non-pulsing XBs~\cite[see][for review]{2012arXiv1206.2727P}.  {We note that burst oscillations have only been detected once in the case of GS~1826-238~\citep{2005ApJ...634.1261T}, despite many bursts having been observed over the course of the RXTE lifetime.  We discuss this source in more detail in \S\ref{sec:RESULTS}.}   The sample was further reduced by observational constraints, such as sources that were too faint (such as MXB~$1659-298$) or are not observed in a hard state.  { At this stage we were left with 8 sources out of 10 LMXBs with coherent pulsations detected during the X-ray bursts but not in their persistent emission~\citep{2012ARA&A..50..609W}, and Aql~X-1 \citep[which once displayed coherent oscillations in its persistent emission over a period of $\approx100$~s,][]{1998ApJ...495L...9Z}. We} include 4U~$1705-44$ from Paper~I for completeness, despite no burst oscillations having been detected from this source, with a view to excluding it for analysis from which we seek to make firm conclusions about spin. We list our final sample, with their pertinent properties in table~\ref{tab:nh}.    We reduce all hard state datasets, which we define as those data possessing a mean hardness ratio $\ge2$ using the intensities measured in the $7.50-18.50$~keV and $4.0-6.0$~keV bands.  We present the final list of observations in tables~\ref{tab:obs}~\&~\ref{tab:obs2}.  

\input{nh.tex}

\section{Data Reduction and Analysis}
\label{sec:red}

\input{obstable.tex}
\input{obstable2.tex}
We extract PCA and HEXTE spectra of each source as detailed in Paper~I, taking care to remove any contamination by X-ray bursts.  For complete consistency, we consider PCA data from PCU2 from each source and HEXTE data from cluster B only.  We fit the spectra with an absorbed Comptonisation model with an additional Gaussian component to model the fluorescent Fe emission.  In the spectral fitting package \xspec, this model takes the form of \totalmod, {within which \compps~models the Comptonised emission and its reflection by material in the accretion disc}.  As this work concerns the nature of NSs exclusively, we adjust the model from the approach in Paper~I such as to have a blackbody-shaped seed photon spectrum, which is the same as the \xspec~model {\sc bbodyrad}.  

\begin{figure}
\begin{center}
\includegraphics[width=0.48\textwidth]{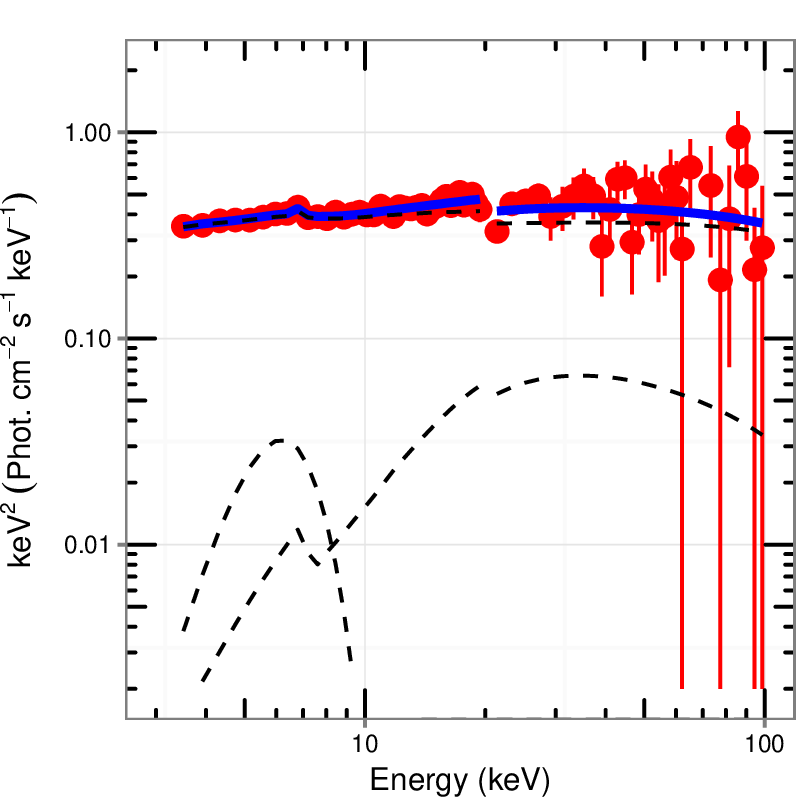}
\end{center}
\caption{Example of unfolded spectral fit (blue line) using data from 4U~$1608-52$ (red). For illustration purposes we show the respective contributions of Comptonised and reflected emission, as well as that of the Gaussian component \label{fig:ufspec}.  For clarity, the emission from the Galactic ridge is omitted.  }
\end{figure}

We model emission from the Galactic ridge (GR) as an additional set of fixed components comprising a $\Gamma=2$ powerlaw and zero-width Gaussian that peaks at an energy of 6.6 keV \citep{2003A&A...410..865R}, the flux of which we calculate using the relation of \cite{2006A&A...452..169R}.  As noted in Paper~I, the close proximity of 4U~$1728-33$ to the Galactic centre means that there should be a substantial contribution of GR emission to these spectra.  In this work we consider an additional two datasets from earlier in the RXTE mission, when the source was less luminous.  On attempting to fit these spectra we noticed clear residual excesses in the spectral model at energies consistent with the line from GR emission, and unusual residuals at lower energies.  This is clearly indicative that we have over-compensated for the GR emission in this source.  To obtain our best estimate of the strength of GR in this instance, we perform a joint fit using all the spectra of 4U~$1728-33$, with the GR flux as the only tied parameter and with the GR spectral shape fixed.  We recover a value of $3.2\times10^{-11}~{\rm erg~cm^{-2}~s^{-1}}$, just over half of our original estimate of $6.15\times10^{-11}~{\rm erg~cm^{-2}~s^{-1}}$, which we use in all subsequent analysis.  { In figure~\ref{fig:ufspec} we show an example best-fit of our model to a spectrum from 4U~1608-52.}

To explore the parameter space we make use of Bayesian X-ray Analysis software \citep[BXA,][]{2014A&A...564A.125B} that connects nested sampling algorithm MultiNest \citep{2009MNRAS.398.1601F} to \xspec.  We bin samples from the posterior distribution produced by BXA for a given spectrum to produce 1D and 2D confidence regions in a given parameter space. We measure the $3.0-200.0$~keV flux for every point in each posterior chain, and calculate the Compton amplification factor $A$ as a ratio of the total luminosity $L_{Tot}$ ($=4\pi f_{3-200}$ $D^2$) to the seed photon luminosity $L_{BB}$ ($=4 \pi R_{BB}^2 \sigma kT_{BB}^4$, where $R_{BB}^2=N (D/\mathrm{10~kpc})^2$).  As both $L_{Tot}$ and $L_{BB}$ are proportional to the distance squared, $A$ is actually  a distance independent quantity.

Finally, to calculate the luminosity as a fraction of Eddington luminosity we take the ratio of the measured $3-200$ keV flux to the peak flux of type-I X-ray bursts (table~\ref{tab:nh}) that exhibit PRE, when a source is presumed to be emitting at the Eddington luminosity $L_{Edd}$. This ratio, which we henceforth refer to as $L/L_{Edd}$, can be considered an excellent proxy for the accretion rate of the source and is a particularly convenient tool for discussing the luminosity of  Galactic sources because it negates the need to involve source distances, which are often subject to various uncertainties.  {We note that for GS~$1826-238$ PRE bursts were not detected during the RXTE era, however, the type-I X-ray bursts from this source had a remarkably consistent profile with an average peak flux $33\times 10^{-9}~{\rm erg~s^{-1}~cm^{-2}}$ \citep{2004ApJ...601..466G}.  A recent investigation by \cite{2016ApJ...818..135C} using NuSTAR data found evidence for PRE in one burst, with peak bolometric flux of $40\pm3 \times 10^{-9}~{\rm erg~s^{-1}~cm^{-2}}$,  however, in the interests in consistency across the source sample we will use the RXTE-era value in our calculation of $L/L_{Edd}$.    }

\subsection{SAX~J1750.8-2900}
\label{sec:1750}

The line-of-sight absorption column to SAX~J1750.8-2900 is not well-known.  The \he~$N_H$ tool, which uses the 21~cm survey of \cite{2005A&A...440..775K}, displays a wide range of values from $0.5-1.6\times10^{22}{\rm~cm^{-2}}$ and presents an average weighted value of $0.927\times10^{22}~{\rm cm^{-2}}$.  Consulting the literature, we find that \cite{2015ApJ...801...10A} discussed the value of $N_H$ in some detail, and performed an analysis of several \emph{Swift} spectra with the value of $N_H$ tied between the fits.  Their procedure obtained $N_H=4.3\times10^{22}{\rm~cm^{-2}}$, which is consistent with the model-dependent extremes of $2.5-6.0\times10^{22}{\rm~cm^{-2}}$ found using BeppoSAX data \citep[][]{1999ApJ...523L..45N} and $4.0-5.9\times10^{22}{\rm~cm^{-2}}$ using XMM data \citep{2012ApJ...749..111L}.  In our analysis we therefore adopt $N_H=4.3\times10^{22}{\rm~cm^{-2}}$.

All hard state observations of this source were made during 2008.  We have concerns about HEXTE data obtained during this epoch.  We analysed two observations of the Crab nebula (93802-02-15-00,93802-02-16-00) taken in the same week as one of the SAX~J1750.8-2900 datasets.  Fitting with the standard absorbed ($N_H=0.38\times10^{22}{\rm~cm^{-2}}$) power law model gave a best fit $\chi^2_\nu=100/82\sim1.23$ and $112/82\sim1.37$ whereas in earlier epochs $\chi^2_\nu$ is typically $<0.9$.  Inspection of the fit residuals in this case shows that the HEXTE data form a slope in the fit residuals, with the model in excess of the data at low energy ($\approx20-40$~keV) and a deficient at higher energies ($\approx50-65$~keV).  We choose not to utilise the HEXTE data during this epoch.  To compensate for the lack of high energy coverage we include PCA data in the range $20.0-45.0$~keV.  Two potential problems with this approach are that the PCA is less-well calibrated at energies $>20$~keV, and that we might not be able to constrain the electron temperature if it is particularly high (i.e. if $kT_e>50$~keV, as in the case of 4U~1608-52).  In terms of calibration, we note the work of \cite{2014ApJ...794...73G}, who computed an epoch-dependent `correction curve' for PCA spectra based on accumulated Crab observations.  For data such as these SAX~J1750.8-2900 observations, from later PCA epochs, the calibration is accurate to within $1\%$ upto $\approx31$~keV (compared to $3-4\%$ in earlier epochs) and can underestimate the flux by $2-4\%$ from $34-45$~keV.  The spectra of    SAX~J1750.8-2900 typically have $20-150$ net counts per channel at energies $>35$~keV, therefore calibration inaccuracies should have a negligible effect on our analysis.

\section{Results}
\label{sec:RESULTS}
\begin{figure*}
\begin{center}
\includegraphics[width=0.9\textwidth]{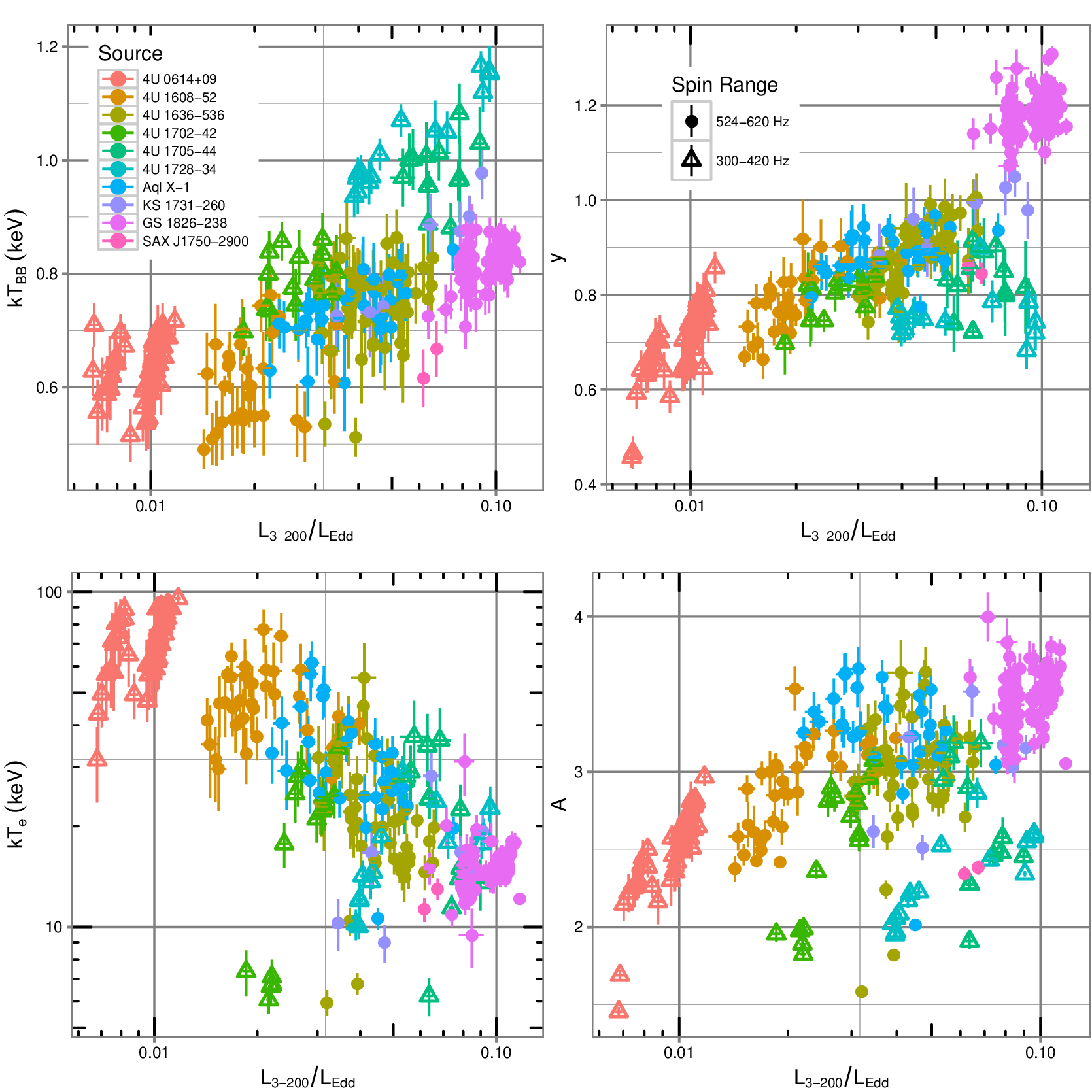}
\end{center}
\caption{Variation of key spectral properties with $L/L_{Edd}$ \label{fig:ktektbb}.}
\end{figure*}

In figure~\ref{fig:ktektbb} we present key spectral properties as a function of $L/L_{Edd}$, with sources denoted by different colours.  Each datum represents the mean posterior estimate of a single spectrum and the associated standard deviation.  To aid our investigation into the effects of NS spin, we divide the sample into two broad spin frequency groups divided about the mean value of our {sample,$\approx500$~Hz}, and in figure~\ref{fig:ktektbb} denote members of each group with circles and triangles respectively. { We note that if the peak bolometric flux measured during a type-I X-ray burst showing signs of PRE from GS~1826-238 is accurate \citep{2016ApJ...818..135C}, the true $L/L_{Edd}$ for this source will be at $\sim82\%$ of the values shown, and emphasise that this change would not be significant enough change to alter our conclusions. }
%% 500 because 510 without 4U1705, 490 with 4U1705

 There are clear trends with $L/L_{Edd}$ in  all four  main parameters characterising the Comptonized spectra.     Formally, the correlations are extremely significant.  For example,  for the Comptonization parameter, the Spearman rank test indicates  a compatibility with the null hypothesis $p_n<2.2\times10^{-16}$ and a correlation of 0.85 between $y$ and $L/L_{Edd}$.  It is also clear that there is significant overlap for each parameter for sources with  different spins. This is further illustrated by the posterior distributions for groups of low- and high-spin sources, figures~\ref{fig:yDIST}, which discussed in detail below. However, figure~\ref{fig:ktektbb} suggests differences in behaviour correlated with spin do exist at a given accretion rate.  Specifically, the low frequency sources all occupy a higher $kT_{BB}$ at a given $L/L_{Edd}$ compared to sources that contain more rapidly rotating NSs.  Likewise, the more slowly rotating NSs appear to have lower $y$ and $A$.

\begin{figure*}
\begin{center}
\includegraphics[height=0.37\textwidth]{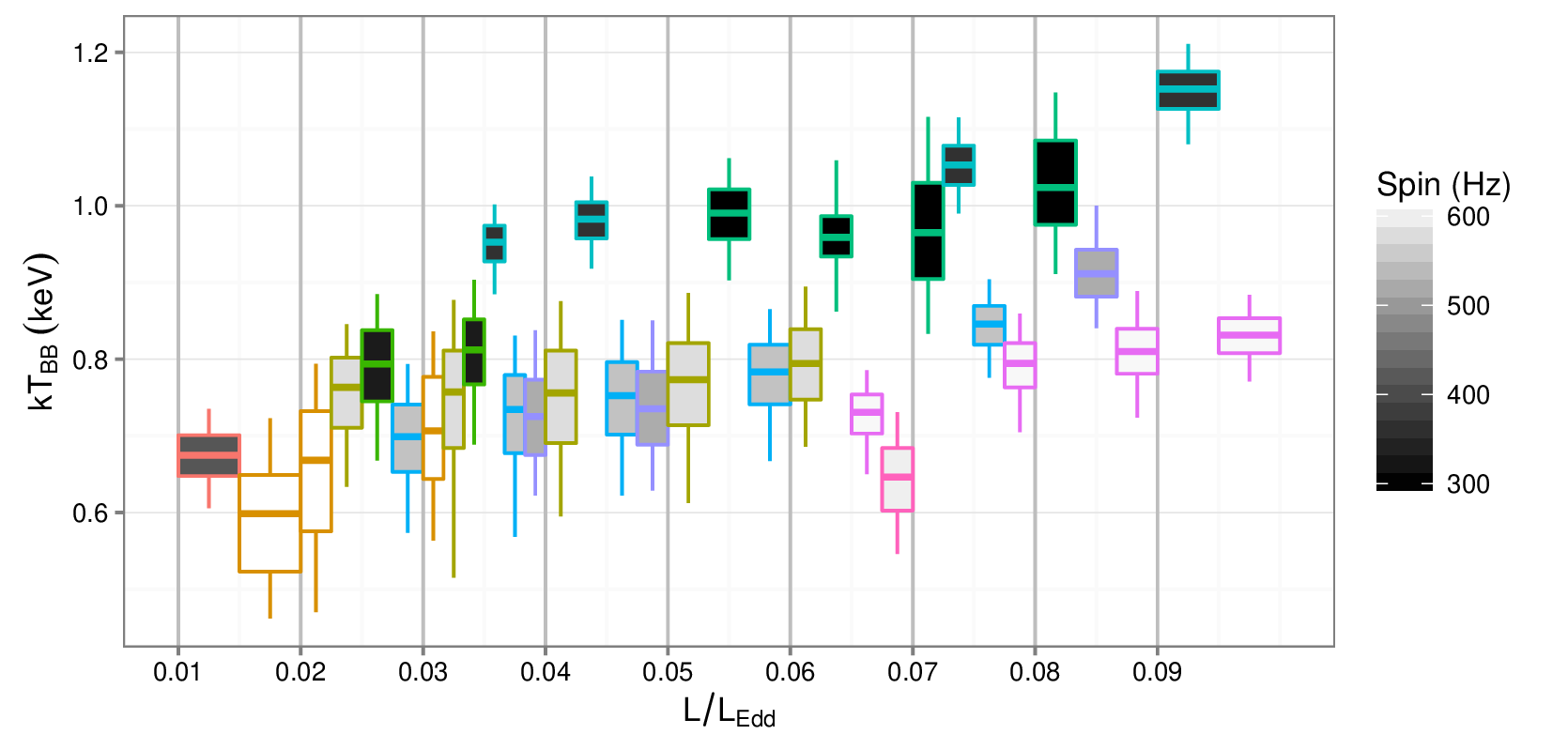}
\includegraphics[height=0.37\textwidth]{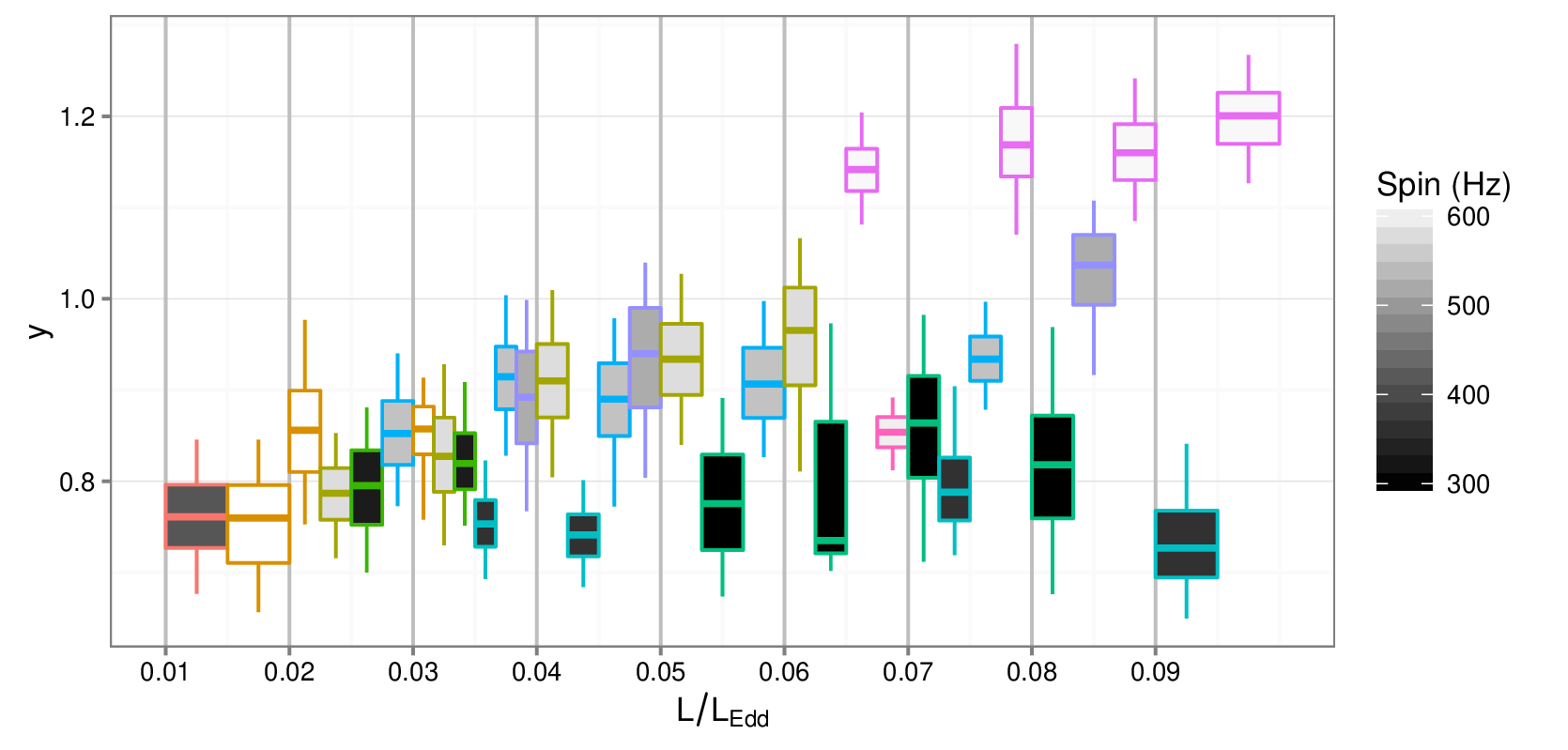}
\includegraphics[height=0.37\textwidth]{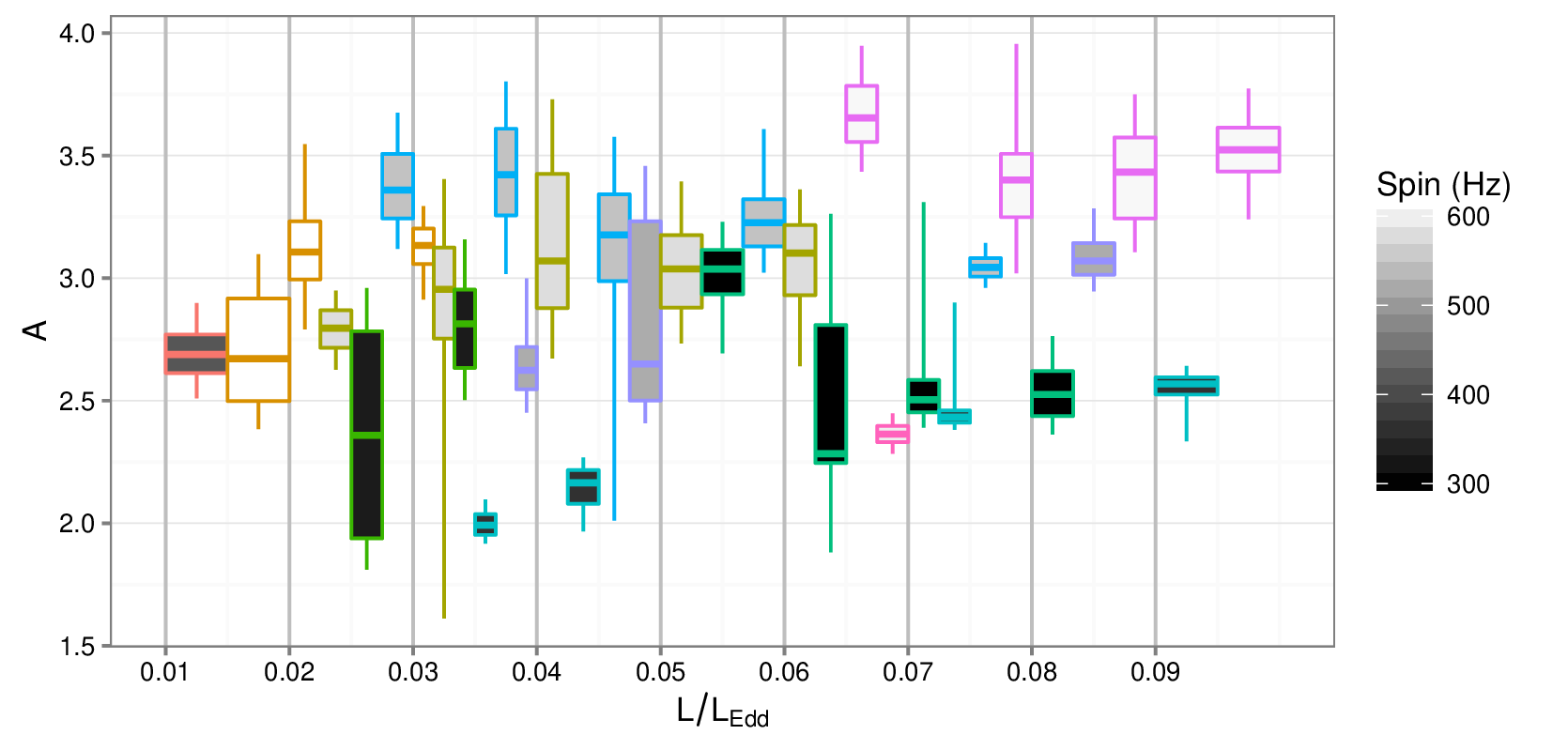}
\end{center}
\caption{Comptonization model parameters binned by $L/L_{Edd}$, with grey vertical lines indicating the bin boundaries.  Each box represents the weighted $25\%-75\%$ interquartile range and each vertical line the $0.05-0.95\%$ bounds for the distribution of that parameter across all posterior chain elements that fall in a given $L/L_{Edd}$ bin for each source, with the line inside the box indicating the position of the median value.  Boxes are offset in $x$ for clarity. \label{fig:fluxres}}
\end{figure*}

\begin{figure*}
\begin{center}
\includegraphics[width=\textwidth]{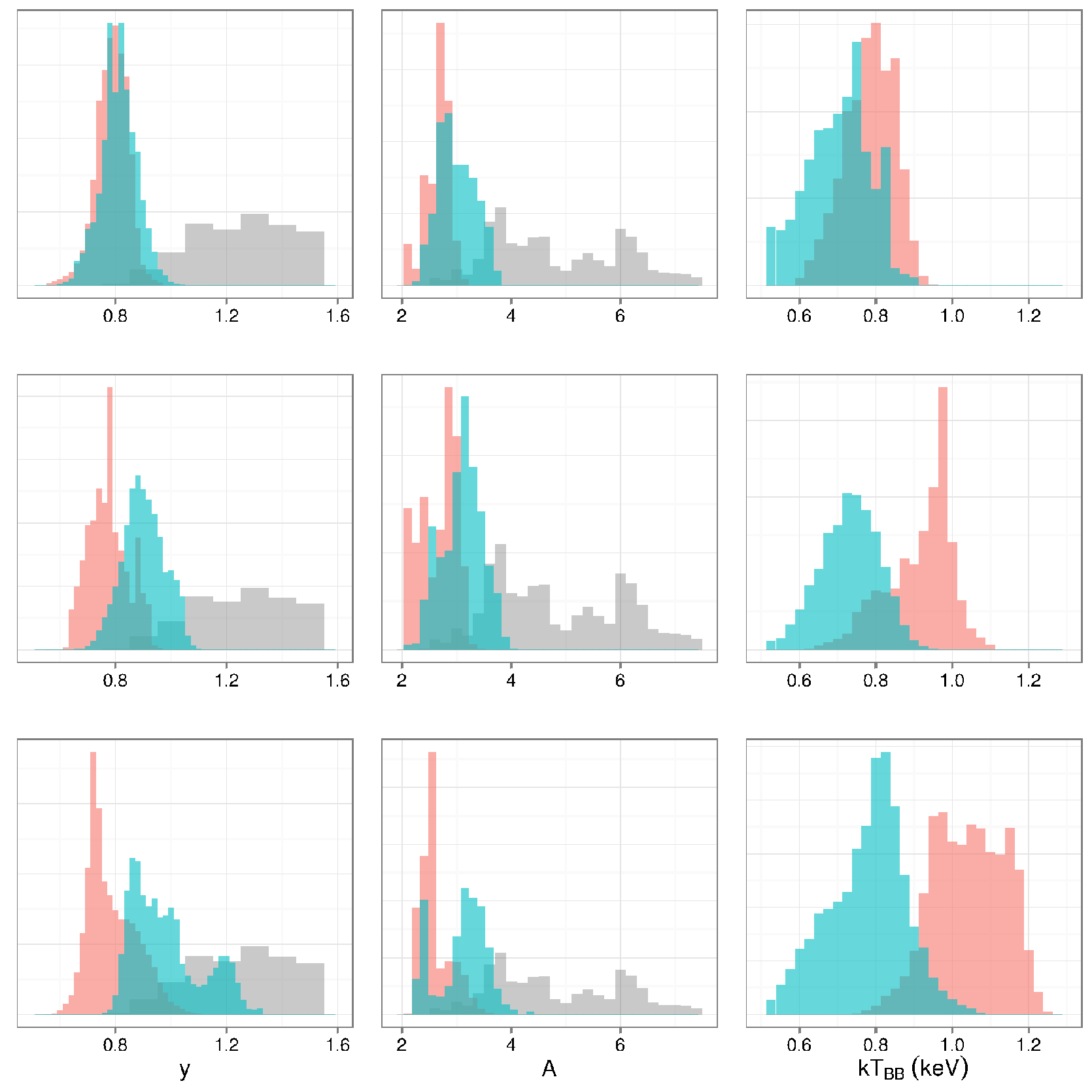}
\end{center}
\caption{Flux-resolved posterior distributions of Compton y-parameter (left), Amplification factor $A$ (center), and $kT_{BB}$ (right) contrasting rapidly spinning NSs ($>500$~Hz, turquoise) and more slowly rotating NSs ($<500$~Hz, pink) with the equivalent distributions for BHs from Paper 1 (grey, left and centre only) \label{fig:bhdists}. { Rows correspond to $L/L_{Edd}=0.015-0.03$, $0.03-0.05$ and $0.05-0.10$ (top-bottom). } BH distribution of $y$ shifted by $+0.05$ (see text) \label{fig:yDIST}.}
\end{figure*}

To test these possible trends connected to source spin we { compute for each source the luminosity resolved mean values  of  spectral parameters and quantiles of their distributions.}  Each element of the posterior chains for every spectrum was allocated a bin based on $L/L_{Edd}$, and then used to compute quantiles for each source if it has a significant presence (equivalent to one entire posterior) in a given flux bin.    In figure~\ref{fig:fluxres} we plot these quantiles ($25\%$,$50\%$ and $75\%$) as a boxplot with each box representing one source for each flux bin, with the spin of each source indicated by the strength of shading.  

Figure~\ref{fig:fluxres} shows that the distributions of $kT_{BB}$ occupy higher values for the three most slowly rotating NSs in all flux bins.  Moreover, these same sources have consistently lower $y$-parameter across the full $L/L_{Edd}$ range, while more rapid rotators have increasingly large values with increasing accretion rate.    The picture in terms of $A$ is less clear cut, with some sources occupying a wider range of values.  However, for a majority of flux bins a slow spinning source will have lower $A$ and $y$ than a more rapidly spinning source.

{ In figure~\ref{fig:bhdists} we present  the one dimensional posterior distributions of $kT_{BB}$, $y$ and $A$ for  low and high-spin sources  in three broad $L/L_{Edd}$ bins. For $y$ and $A$ these distributions are compared with the corresponding distributions for black holes from Paper I. These distributions complement figures \ref{fig:ktektbb} and \ref{fig:fluxres} in illustrating the  dichotomy between low and high spin sources and their trends with the Eddington ratio. These will be further discussed in section 5.}

We ask what is the probability that $kT_{BB}$ will correctly rank the sources by spin when temperature is drawn from a uniform distribution for fixed $L/L_{Edd}$?  
Considering all 10 sources, we can broadly consider to fall into three groups within which the sources overlap.  This makes groups of two groups of three sources, and one group of four sources.  What is the probability that  that $kT_{BB}$ can rank the spectra into such groups by chance?  Assuming $L/L_{Edd}$ is fixed, we generate 10 random values of $kT_{BB}$ from a uniform distribution over the observed range of $kT_{BB}$ at that point ($\approx 0.6-0.85$~keV), and repeat for a large number of iterations.  We then ask how many times are the first three numbers lower than the next 4 numbers, and at the same time those numbers are lower than the final 3 numbers?  This analysis suggests a uniform distribution would produce this ranking result {by chance with the probability of}  $\approx 2.4\times10^{-4}$.

{ As mentioned in section 2,  in the case of GS~$1826-238$ burst oscillations have only been reported once by \cite{2005ApJ...634.1261T}  who analysed  three RXTE observations that had been simultaneously observed with \emph{Chandra} in their search. Their detection is unconfirmed in many other bursts from this source observed by RXTE.  In particular, \cite{2012ARA&A..50..609W} suggested that the authors underestimated the number of search trials in arriving at their significance estimate. Given these considerations we assume that the burst frequency is unknown, so as to test the impact on our results of withdrawing GS 1826-238 from the sample (just as we excluded many other NS LMXBs with unknown or unconfirmed spins). Withdrawing this source does not affect our conclusion that the Comptonizing properties are a function of both the accretion rate and spin. More specifically, removing this source from fig.3 (i.e. the four purple boxes at $L/L_{Edd}~0.65-0.95$) does not affect the split between higher and lower spin sources. In fig.4, removing this source will only affect the bottom row of panels corresponding to highest L/Ledd, with the only noticeable effect being the disappearance of the second peak of the  turquoise histogram in the bottom left panel. This does not affect the existence of the bimodality in these distributions, neither does it change the false alarm probabilities computed above. }

\section{Discussion and conclusions}

We have performed a systematic study of a sample of NS LMXB hard state spectra to investigate possible trends between the spectral properties and the NS rotation frequency.  {  To this end, we modelled} the hard state spectra as Comptonization of blackbody seed photons { on thermal electrons,} and explored the available parameter space for this model using nested sampling.  We note that our approach does not account for the effects of inclination (as the emission will be locally anisotropic), {second-order reflection effects} in both the disc and boundary layer caused by the photons from each other, or relativistic effects from the vicinity of the NS~\citep[see discussion in][]{1985MNRAS.217..291L}.

{ Overall, spectral properties depend strongly on  the Eddington luminosity ratio, with different sources and and different observation of the same source following same dependence, albeit with large scatter. In addition, at the given Eddington luminosity ratio (i.e. same mass accretion rate), there is a difference in spectral properties  between different sources correlated with the NS rotation frequency. These differences are most apparent in terms of the seed photon temperature $kT_{BB}$ and in the behaviour of the Compton $y-$parameter and Compton amplification factor $A$.  Specifically, we find that  at a given accretion rate the seed photon temperature is lower and Compton y-parameter and $A$ are higher for more rapidly rotating NSs  (figure \ref{fig:fluxres}). We do not find such a clear correlation with the NS spin in the case of the electron temperature $kT_e$.}

{ For the standard Shakura-Sunyaev accretion  disc around a slowly rotating NS, more than a half of the energy possessed by infalling material is released on the stellar surface \citep{1988AdSpR...8..135S,2000AstL...26..699S}, in a narrow boundary or spreading layer }\citep{1988AdSpR...8..135S,1999AstL...25..269I,2010AstL...36..848I} where the material decelerates between the Keplarian velocity of the disc and the rotational velocity of the NS surface.  Analysis of Fourier-frequency-resolved spectra of multiple sources by \cite{2003A&A...410..217G} { and \cite{2006A&A...453..253R}} suggests that {  in the soft spectral state,} the boundary layer has an approximately blackbody spectrum of common characteristic temperature $\approx2.4$~keV.   While no such { spectral component} has been identified in the NS hard state spectra, {   some fraction of the energy of the accreting matter is still expected to be released on the surface of the neutron star, in some form of boundary layer}, but its emission is presumably intercepted by the Comptonizing material along the line-of-sight.  Paper~I showed that Comptonisation is generally weaker (smaller $A$ \& $y$) in NS systems than for their BH cousins and this can be understood in terms of additional seed photons { emitted by the neutron star} instigating greater energy losses of hot electrons  and reducing the Comptonizing strength of the corona.  { In Paper~I we estimated that about $\sim 1/3-1/2$ of the accretion energy must be released on the surface of the neutron star in order to explain the difference in the parameters of the Comptonized spectra in BH and NS systems.}

{ \cite{2003A&A...410..217G} and \cite{2006A&A...453..253R} found that in the soft spectral state, the temperature of the spreading layer emission was of the order of the Eddington temperature\footnote{$\sim 2.5$ keV for the distant observer, including colour correction, see \cite{2014PhyU...57..377G} for details.}  and did not depend on the mass accretion rate, which in their sample ranged from $\approx (0.1-1)\times L_{Edd}$. This result supported the theoretical picture of the Eddington flux-limited  levitating  spreading layer supported by the radiation pressure, proposed earlier by \citet{1999AstL...25..269I}. In their model, the spectral shape of the boundary layer emission is determined only by the neutron star compactness \citep[see][for detailed calculation]{2006MNRAS.369.2036S} and its luminosity is regulated by the surface area of the spreading layer.  In the present work we find that in the hard spectral state on the contrary, the seed photon temperature is significantly smaller that the local Eddington value and varies with  X-ray  luminosity of the source, i.e. with the mass accretion rate. This points at the difference in the structure and dynamics of the boundary layer in the soft and hard spectral states.
}

{
The greater the differential between the  linear velocity of the accreting material  and that of the NS surface, the more energy is liberated at the boundary layer.   In the Newtonian approximation the luminosity of the boundary layer $L_{bl}$ is given by equation~\ref{eq:lbl}, which can be expressed in terms of the respective rotation frequencies $f_K$ and $f_{NS}$,
\begin{equation}
L_{bl}=2\pi^2\dot{M}R^2 \left(\alpha f_K- f_{NS}\right)^2.
\label{eq:lbl2}
\end{equation}
One can see from these equations that the luminosity of seed photon produced by the NS surface anti-correlates with the NS spin. For higher spin sources supply of the seed photons will be reduced resulting in stronger Comptonization, in particular in large Compton amplification factor $A$ and Comptonization parameter $y$, in the same way as the dichotomy between BH and NS sources had been explained in Paper I.

Furthermore, assuming the emission from the boundary layer is  well-approximated by a blackbody spectrum, then $L_{bl}\propto T^4$.  This means that if all other factors  (especially the emitting surface area) are similar across the population of NSs, the temperature of the seed photons $kT_{BB} \propto (\alpha f_K - f_{NS})^\frac{1}{2} $.  The maximal Keplerian frequency in the accretion disc is of the order $f_K\sim 1.6-1.8$ kHz \citep{2000AstL...26..699S}, while NS spins vary in the $f_{NS}\sim 0.3-0.6$ kHz range. At $L_X/L_{Edd}\sim 0.05$, the seed photon temperature varies from $kT_{BB}\sim 0.8-1.0$ keV, i.e. by a factor of  about $\sim 1.2-1.3$, between low- and high-spin sources (Fig. \ref{fig:fluxres}). In order to explain such a difference between $f_{NS}\sim0.3$ kHz and $f_{NS}\sim 0.6$ kHz sources, $\alpha\sim 0.7$ is required, which is broadly consistent with the conclusions of Paper I that the accreting material reaches the surface of the neutron star still possessing about $\sim 1/3-1/2$ of its initial  energy.}

\subsection{Can fast spinning neutron stars mimic black holes?}
\label{sec:nsbh}

\begin{figure}
\begin{center}
\includegraphics[width=0.48\textwidth]{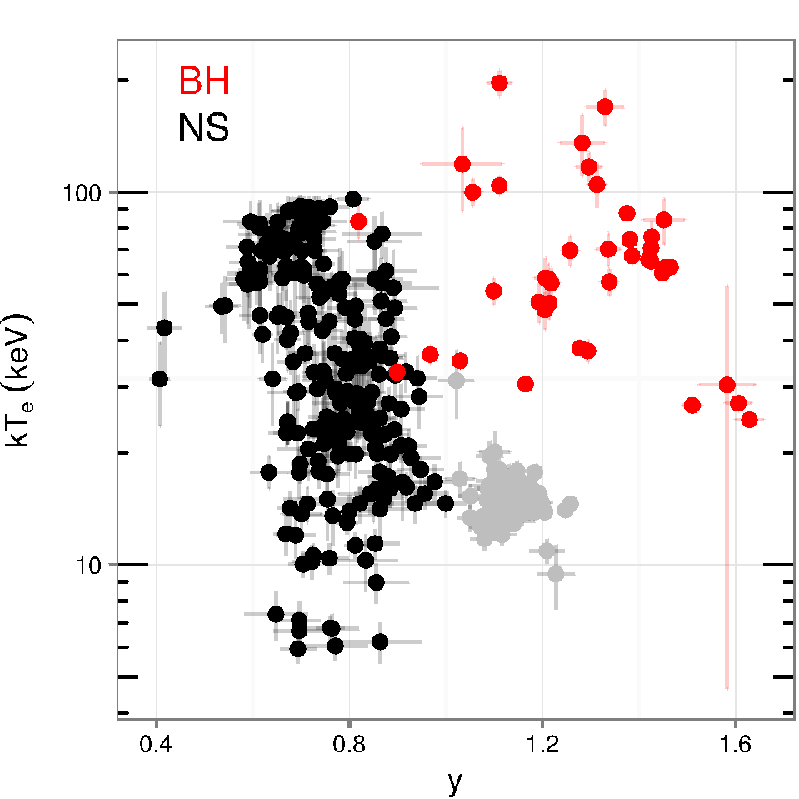}
\end{center}
\caption{Compton y-parameter and $kT_{e}$ for NS spectra from this work (black) and BH spectra from paper~I (red).  Points corresponding to GS~$1826-238$ are highlighted in grey (see discussion in \S\ref{sec:nsbh}). 
\label{fig:ktey}}
\end{figure}

{ From equation~\ref{eq:lbl2}, at sufficiently high NS spin frequencies the luminosity of the boundary layer can become small. Theoretically, this can make accreting neutron stars similar to black holes from the point of view  of the energy balance in the Comptonization region. Indeed, such a behaviour is  evident in Figure~\ref{fig:yDIST} where the high-spin sources at high Eddington ratios progressively overlap with the region of Comptonization parameters, identified with black holes in Paper I. This dilutes the boundary between black holes and neutron stars. From Figure~\ref{fig:ktektbb} one can see that three sources expand into  $y\gtrapprox1.0$ regions at higher $L/L_{Edd}$: 4U~$1636-536$, KS~$1731-260$ and, most significantly, GS~$1826-238$.  This behaviour is strongest at the highest Eddington ratios (but still compatible with the source being in the hard spectral state).  All these sources have high spin  ($>500$~Hz), and their behaviour markedly contrasts with that of the slower sources at similar $L/L_{Edd}$, such as 4U~$1728-34$, the Comptonizing strength of which remains low. 
For GS~$1826-238$, which has one of the highest spins and mass accretion rates in our sample, the Comptonization parameter can reach values $y\sim 1.25$, which is inside the region occupied by BH LMXBs.\footnote{We note that in the strictest sense these $y$ should not be compared directly; the different seed photon model used in the current work means that we are in effect comparing two different models, however, experimentation by fitting the spectra in this work with both models shows that $y$ is fairly resilient between both approaches, and on average $y$ from using the blackbody seed photon spectrum is $\approx0.05$ higher than that recovered from fitting using a multicolour disc blackbody seed spectrum.}   However, such high values of $y$ are achieved only at high Eddington ratios, $L_X/L_{Edd}\sim 0.1$, at which neutron stars are characterised by extremely  low electron temperature, $kT_e\sim 10-20$ keV (Figure~\ref{fig:ktektbb}), well below values typical for black holes (Fig.~5 in Paper I). This behaviour in $kT_e-y$ distinguishes NSs from black holes and is illustrated in figure~\ref{fig:ktey}.  A  detailed comparison of high spin neutron stars with black holes  will be a subject of a separate study. { As a final point, we mention that withdrawing GS~$1826-238$ from the sample (see discussion in \S4) does not  negate the encroachment of the more luminous and high-spin sources onto the territory of Comptonization strength occupied by BHs (Fig.~\ref{fig:yDIST}), but makes it less extreme. This is illustrated in figure~\ref{fig:ktey}, where the GS~$1826-238$ points are shown in grey.     
}

\section*{Acknowledgments}
Our sincere thanks to the anonymous referee, whose comments have improved the clarity of the paper.  We acknowledge the use of the Legacy Archive for Microwave Background Data Analysis (LAMBDA), part of the High Energy Astrophysics Science Archive Center (HEASARC). HEASARC/LAMBDA is a service of the Astrophysics Science Division at the NASA Goddard Space Flight Center.  MG acknowledges hospitality of the Kazan Federal University (KFU) and support by the Russian Government Program of Competitive Growth of KFU and RFBR grant No. 15-42-02573.

\bibliography{spin}{}
\bibliographystyle{mn2e}

\bsp

\label{lastpage}

\end{document}

%% file: newcom.tex
\newcommand{\nm}{\mbox{\ensuremath{\mathrm{~\nm}}}}

%% file: nh.tex
\begin{table}
  \begin{tabular}{lccc}
Source & ${\rm N_H}$ & Spin &  $F_{peak}$  \\
 & $\rm{10^{22}~cm^{-2}}$ & Hz  &  ${\rm 10^{-9}~erg~s^{-1}~cm^{-2}}$ \\
\hline 
4U~1608-52 & 1.81 & 620 & 132   \\
GS~1826-238 & 0.171 & 611 & 33 \\
SAX~J1750.8-2900 & $4.3^{*}$ & 601 & 49.2  \\
4U~1636-536 &  0.27 & 581 & 64    \\ 
Aql~X-1 & 0.28 & 550 &  89  \\ 
KS~1731-260 & 0.31 & 524 & 43    \\
4U~0614+09 & 0.448 & 415  & 200 \\
4U~1728-33 & 1.24  & 363 & 84  \\ 
4U~1702-429 & 1.12 & 329 & 76  \\
4U~1705-44 & 0.67  & $298^{**}$ & 39.3   \\ 
  \end{tabular}
\caption{Source Properties.   For our spectral analysis we use equivalent Hydrogen column densities $N_H$ inferred from 21 cm emission \citep{2005A&A...440..775K}, except in the case of SAX~J1750.8-2900 ($^*$see \S~\ref{sec:1750}).  The mean peak fluxes $F_{peak}$ of X-ray bursts are those reported by \citet{2008ApJS..179..360G}  except for GS~1826-238 for which we use figures reported by \citet{2004ApJ...601..466G} (see \S~\ref{sec:red}) and for 4U~0614+09 \citep{2010A&A...514A..65K}. Spin frequencies correspond to burst oscillation frequencies presented by \citet{2008ApJS..179..360G} and \citet{2008ApJ...672L..37S} (in the case of 4U~0614+09).  $^{**}$For 4U~$1705-44$ we quote the difference in kHz QPO frequencies as an approximation of the true spin frequency \citep{1998ApJ...498L..41F}, baring in mind the possible uncertainties inherent to this method of spin determination \citep{2014AN....335..168W}.}
 \label{tab:nh}
\end{table}

%% file: obstable.tex
\begin{table*}
  \begin{tabular}{lrl}
{\bf Source} & {\bf Proposal ID} & {\bf Observation Suffix} \\
\hline 
4U~0614+09 & 50031  &  -01-01-06  -01-02-00  -01-02-02  -01-02-02  -01-02-02  -01-02-04  -01-02-04  -01-02-05  -01-02-05  -01-02-05  \\ & &  -01-02-07  -01-05-00  -01-05-00  -01-05-00  -01-05-00  -01-05-01  -01-05-02  -01-05-02  -01-04-01  \\ & &  -01-04-02  -01-04-03  -01-04-07  -01-04-04  -01-04-14  -01-04-15  -01-04-05  -01-04-06  -01-04-06  \\ & &  -01-04-06  -01-04-08  -01-04-09  -01-04-09  -01-04-00  -01-04-00  -01-04-00  -01-04-00  -01-04-10  \\ & &  -01-04-11  -01-04-16  -01-04-17  -01-04-18  -01-04-12  -01-04-13  -01-04-13  -01-04-13  \\
 & 60424  &  -01-02-00  -01-02-00  -01-02-01  \\ 
 & 70014  &  -04-01-00  -04-01-00  -04-01-01  \\ 
 & 70015  &  -03-01-00  -03-01-00  -03-01-01  -03-02-00  \\ 
 & 80037  &  -01-12-00  \\ 
 & 80414  &  -01-01-01  -01-01-01  -01-03-00  \\ 
 & 90422  &  -01-03-00  \\ 
 & 92411  &  -01-09-00  \\ & & \\
4U~1608-52 & 10094  &  -01-07-00  -01-07-00  \\ 
 & 40437  &  -01-02-00  \\ 
 & 50052  &  -01-33-00  -01-37-00  \\ 
 & 60052 & 03-01-06 03-02-02 03-02-04 03-02-06 \\
 & 70058  &  -01-01-00  -01-02-00  -01-04-00  -01-04-01  -01-05-00  -01-06-00  -01-10-00  -01-11-00  -01-12-00  -01-13-00  \\ 
 & 91405  &  -01-14-01  -01-15-01  -01-16-00  -01-18-00  -01-20-00  -01-24-01  -01-25-00  -01-25-01  -01-26-00  -01-26-02  \\ & &  -01-28-00  -01-28-01  -01-29-01  -01-30-00  -01-35-00  -01-35-01  -01-38-01  -01-39-00  -01-39-01  \\ & &  -01-40-00  -01-40-01  -01-40-03  \\  & & \\
4U~1636-536 & 90409  &  -01-01-00  \\ 
 & 91024  &  -01-06-00  -01-07-00  -01-08-00  -01-09-00  -01-26-00  -01-27-00  -01-44-00  -01-51-00  -01-52-00  -01-72-00  \\ & &  -01-73-00  -01-73-00  -01-74-00  -01-75-00  -01-76-00  -01-77-00  -01-97-00  -01-21-10  -01-22-10  \\ & &  -01-23-10  -01-24-10  -01-25-10  -01-47-10  -01-48-10  -01-50-10  -01-68-10  -01-72-10  -01-11-00  \\ & &  -01-78-01  \\ 
 & 92023  &  -01-06-00  -01-06-01  -02-08-00  -01-08-00  -02-09-00  -02-10-00  -02-11-00  -01-11-00  -02-12-00  -01-12-00  \\ & &  -01-30-00  -01-31-00  -01-32-00  -01-33-00  -01-34-00  -01-35-00  -01-56-00  -01-57-00  -01-58-00  \\ & &  -01-59-00  -01-97-00  -01-98-00  -01-01-10  -01-02-10  -01-20-10  -01-46-10  -01-50-10  -01-05-20  \\ & &  -01-14-20  -01-15-20  -01-17-20  -01-18-20  -01-86-10  -01-49-10  -01-88-10  \\ & &  \\ 
4U~1702-429 & 40025  &  -04-03-01  -04-02-00  \\  
 & 50030  &  -01-05-00  -01-10-02  -01-10-02  -01-10-03  -01-10-03  -01-13-01  -01-13-02  -01-13-02  \\ 
 & 80033  &  -01-01-01  -01-01-01  -01-21-02  -01-21-02  -01-20-01  -01-20-01  -01-20-04  -01-20-04  \\ & & \\ 
4U~1705-44 & 20073  &  -04-01-00  \\ 
 & 40034  &  -01-08-02  -01-09-04  -01-09-01  -01-09-07  \\ 
 & 40051  &  -03-10-00  -03-12-00  -03-12-00  -03-13-00  \\ 
 & 91039  &  -01-01-41  -01-01-42  -01-01-43  -01-01-50  -01-01-51  -01-01-61  \\ & &  \\ 
  \end{tabular}
  \caption{RXTE Observations 1}

  \label{tab:obs}
\end{table*}

%% file: obstable2.tex
\begin{table*}
  \begin{tabular}{lrl}
{\bf Source} & {\bf Proposal ID}  & {\bf Observation Suffix} \\
\hline 
4U~1728-33 & 10073  &  -01-10-00  \\ 
 & 40027  &  -06-01-00  -06-01-02  -06-01-06  -06-01-03  -06-01-03  -06-01-03  -06-01-08  -08-02-01  -06-01-01  -06-01-01  \\ & &  -06-01-02  -06-01-04  -06-01-05  \\ 
 & 91023  &  -01-06-00  -01-06-00  \\ 
 & 92023  &  -03-47-10  -03-48-00  -03-49-00  -03-67-00  -03-69-00  -03-66-10  -03-83-10  \\ & &  \\ 
Aql~X-1 & 20092  &  -01-01-02  -01-02-02  \\ 
 & 40033  &  -10-01-00  -10-01-01  -10-01-01  -10-01-02  -10-01-02  -10-01-02  \\
 & 40048  &  -01-03-00  -01-03-00  -01-03-00  -01-03-00  -01-07-00  -01-07-00  -01-07-00  \\ 
 & 40049  &  -01-02-00  -01-02-01  -01-03-00  -01-03-00  -01-03-00  \\ 
 & 40432  &  -01-02-00  -01-02-00  -01-03-00  \\ 
 & 50049  &  -01-04-00  -01-04-01  -02-03-01  \\ 
 & 70426  &  -01-01-00  \\ 
 & 90017  &  -01-01-00  -01-01-02  -01-02-00  -01-02-01  -01-03-01  -01-04-00  -01-06-00  -01-06-01  -01-07-00  -01-08-00  \\ & &  -01-08-01  -01-08-02  -01-09-00  -01-09-02  \\ 
 & 91028  &  -01-01-00  -01-03-00  -01-04-00  -01-05-00  -01-07-00  \\ & &  \\ 
KS~1731-260 & 40025  &  -02-03-02  -02-04-00  -02-04-01  -02-05-00  -02-05-01  -02-02-00  -02-03-01  \\
 & 50031  &  -02-01-00  -02-01-00  -02-01-00  -02-01-07  -02-01-09  -02-01-10  -02-01-11  -02-01-12  -02-01-01  -02-01-01  \\ & &  -02-01-01  -02-01-02  -02-01-02  -02-01-02  -02-01-02  -02-01-03  -02-01-03  -02-01-03  -02-01-04  \\ & &  -02-01-04  -02-01-05  -02-01-05  -02-01-05  -02-01-08  -02-02-04  -02-02-05  -02-02-05  -02-02-00  \\ & &  \\ 
SAX~J1750.8-2900 & 93432  &  -01-05-02  -01-05-04  -01-06-00  \\ & & \\
GS~1826-238 & 30060  &  -03-03-00  -03-03-00  -03-03-00  \\ 
 & 50035  &  -01-01-00  -01-01-00  -01-01-02  -01-01-02  -01-01-02  \\ 
 & 70025  &  -01-01-00  -01-01-00  -01-01-00  -01-01-00  \\ 
 & 70044  &  -01-01-00  -01-01-00  -01-01-00  -01-01-05  -01-01-01  -01-01-02  -01-02-00  -01-02-02  -01-02-03  -01-03-00  \\ & &  -01-04-00  -01-04-00  -01-04-00  \\ 
 & 80048  &  -01-01-02  -01-01-17  -01-01-11  -01-01-00  -01-01-00  -01-01-00  -01-01-01  -01-01-01  -01-01-13  -01-01-03  \\ & &  -01-01-03  -01-01-14  -01-01-05  -01-01-04  -01-01-04  -01-01-04  -01-01-07  -01-01-07  \\ 
 & 80049  &  -01-01-00  -01-03-00  -01-03-00  -01-03-00  -01-03-01  -01-03-01  -01-03-01  -01-03-02  -01-03-02  -01-03-03  \\ & &  -01-03-03  -01-04-00  \\ 
 & 80105  &  -11-01-00  -11-01-00  \\ 
 & 90043  &  -01-01-00  -01-01-00  -01-01-00  -01-01-04  -01-01-02  -01-01-02  -01-01-01  -01-01-01  -01-01-01  -01-01-01  \\ & &  -01-01-03  -01-02-01  -01-02-02  -01-02-03  -01-03-00  -01-03-01  -01-03-01  \\ 
 & 91017  &  -01-01-02  -01-01-01  -01-01-01  -01-01-01  -01-01-01  -01-01-00  -01-02-17  -01-02-05  -01-02-04  -01-02-00  \\ & &  -01-02-12  -01-02-13  -01-02-14  -01-02-01  -01-02-07  -01-02-06  -01-02-08  -01-02-10  \\
 & 92031  &  -01-01-00  -01-01-02  -01-01-03  -01-01-04  -01-01-06  -01-01-06  -01-01-01  -01-01-07  -01-01-08  -01-01-09  \\ & &  -01-02-01  -01-02-05  -01-02-04  -01-02-03  -01-02-02  -01-02-00  -01-02-00  -01-02-00  -01-02-00  \\ 
 & 92703  &  -01-01-01  -01-02-01  -01-02-02  -01-02-03  -01-02-03  -01-03-01  -01-03-00  -01-03-03  -01-03-02  -01-04-00  \\ & &  -01-04-01  -01-04-02  \\  
  \end{tabular}
  \caption{RXTE Observations 2}

  \label{tab:obs2}
\end{table*}